\documentclass[%
 reprint, showpacs,
 amsmath,amssymb,
 prl,
]{revtex4-1}

\usepackage{graphicx}
\usepackage{dcolumn}
\usepackage{bm}

\newcommand{\Ba}{Ba$^+$ }

\begin{document}

\preprint{APS/123-QED}

\title{Neutral atom wavelength compatible 780 nm single photons from a trapped ion via quantum frequency conversion}

\author{James D. Siverns$^1$, John Hannegan$^1$ and Qudsia Quraishi*$^{1,2}$}
\vspace{5mm}
\affiliation{%
 $^1$Joint Quantum Institute, University of Maryland, College Park, MD 20742\\
 $^2$Army Research Laboratory, 2800 Powder Mill Rd., Adelphi, MD 20783\\ 
*corresponding author: quraishi@umd.edu
}%

\date{\today}

\begin{abstract}

\end{abstract}

\maketitle
\textbf{Quantum networks consisting of quantum memories and photonic interconnects can be used for entanglement distribution \cite{Duan03,Kimble08}, quantum teleportation \cite{Pirandola15} and distributed quantum computing \cite{Spiller06}. Remotely connected two-node networks have been demonstrated using memories of the same type: trapped ion systems \cite{Hucul15}, quantum dots \cite{Gao2015} and nitrogen vacancy centers \cite{Gao2015,Hensen2015}. Hybrid systems constrained by the need to use photons with the native emission wavelength of the memory, have been demonstrated between a trapped ion and quantum dot \cite{Meyer15} and between a single neutral atom and a Bose-Einstein Condensate \cite{Lettner11}. Most quantum systems operate at disparate and incompatible wavelengths to each other so such two-node systems have never been demonstrated. Here, we use a trapped 138\Ba ion and a periodically poled lithium niobate (PPLN) waveguide, with a fiber coupled output, to demonstrate 19\% end-to-end efficient quantum frequency conversion (QFC) of single photons from 493 nm to 780 nm. At the optimal signal-to-noise operational parameter, we use fluorescence of the ion to produce light resonant with the $^{87}$Rb $D_2$ transition. To demonstrate the quantum nature of both the unconverted 493 nm photons and the converted photons near 780 nm, we observe strong quantum statics in their respective second order intensity correlations. This work extends the range of intra-lab networking between ions and networking and communication between disparate quantum memories.}

A quantum network may be established by interfering photons emitted by quantum memories. Connecting different types of quantum memories for hybrid networking requires overcoming the disparate photon wavelengths emitted by each quantum memory. Given advances in modularity in trapped ion and neutral atom architectures, a hybrid system with modular inter-connectivity is advantageous.  In the case of photons emitted from trapped ions (with fiber attenuations of 70 dB/km at 369 nm for Yb$^+$ and 50 dB/km at 493 nm for Ba$^+$) generating photons in, or converting photons to, the near-infra-red range (with a fiber attenuation of 3.5 dB/km) would significantly extend the networking range between trapped ions and provide the ability to match the wavelength of another quantum memory. 

Trapped ions \cite{Siverns2017-2} are an excellent candidate for elementary logical units \cite{Monroe14} of a network as many pre-requisite components have been shown, including: modularity for photon generation and detection \cite{Hucul15, Brown2016,Monroe14}, quantum computation \cite{Bermudez17,Linke2017} and excellent single photon emission properties \cite{Blatt17}. In this work, we overcome a challenge to extending the networking range of trapped ions by frequency converting the ion light. We demonstrate the conversion of 493 nm photons, emitted from a single barium ion, to a 780 nm wavelength resonant with the $D_2$ transition in neutral $^{87}$Rb. This conversion provides the two-fold benefit of paving the way for neutral-ion hybrid networking and communication as well as extending the networking capability of barium ions from 100s of meters to several kilometres allowing for both an ion-ion and neutral-ion local quantum intranet \cite{Siverns17}. 

\begin{figure*}
	\centering
	\includegraphics[width=\textwidth]{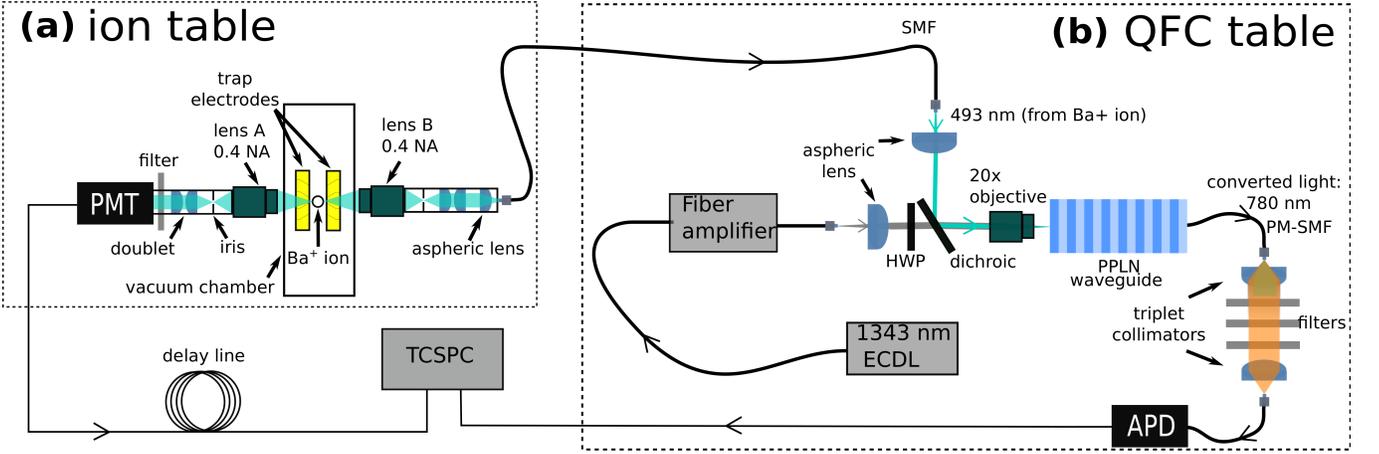}
	\caption{(a) Schematic of trapped ion setup showing the two separate optical collection paths which are situated on opposite sides of an ultra-high vacuum chamber. The vacuum windows are anti-reflection coated and allow optical access for two 0.4 NA objectives, lens A and lens B. The 493 nm photons collected by lens B are collimated and launched into single mode fiber (SMF) of a few meters in length. A photomultiplier tube (PMT) measures photons collected by lens A. (b) Shows the quantum frequency conversion (QFC) setup including a dichroic mirror that combines 493 nm and pump photons. A 20x objective couples the combined light into a periodically poled lithium niobate (PPLN) waveguide with a small length of polarization maintaining single mode fiber (PM-SMF) butt-coupled to its output facet. The light emitted from this fiber is free-space propagated to allow for use of optical filters and then recoupled into a standard SMF, a few meters in length, for detection by an avalanche photodiode (APD). Photo detection events are monitored with a time-correlated two-channel single photon counter (TCSPC).}
	\label{Fig:experiment}
\end{figure*}

Quantum frequency conversion (QFC) \cite{Kumar90} in a periodically poled lithium niobate (PPLN) waveguide, frequency converts a photon of one frequency to another while preserving its quantum properties \cite{lenhard17,Ates12,Ikuta2011}. PPLNs provide a modular component for integration in a quantum network. Using a three-wave parametric $\chi^{(2)}$ nonlinearity, QFC generates an output photon via optical difference frequency generation (DFG) between a pump and input photon. By selecting a pump frequency lower than the desired output frequency, we can minimize noise photons at undesired optical frequencies resulting from spontaneous parametric downconversion and anti-Stokes Raman processes \cite{Pelc10}. 
 
As shown in Fig.~\ref{Fig:experiment}(a), a single $^{138}$Ba$^+$ ion is trapped using electrodes in a blade trap configuration which is housed in an ultra-high vacuum chamber on the ion table \cite{Siverns17}. Windows in the vacuum chamber allow for two collection lenses, each with a numerical aperture (NA) of 0.4, to collect ion fluorescence from either side of the chamber simultaneously. The windows are anti-reflection coated and the total fluorescence collection is $\sim$8$\%$ at 493 nm. Collection lens A directs free-space light into a photomultiplier tube (PMT), with a specified quantum efficiency (QE) of around 6$\%$ (Hamamatsu: H11870-01 \cite{disclaimer}) at 493 nm, via a doublet lens, 493 nm notch filter and an iris. The $^{138}$Ba$^+$ ion is Doppler cooled with 493 nm light on the $^2$S$_{1/2}$ - $^2$P$_{1/2}$ transition and re-pumped out of the meta-stable $^2$D$_{3/2}$ state with light near 650 nm. 

The Doppler and repump beam frequencies are stabilized to an optical wavemeter to $\sim$2 MHz. With this setup, we observe a maximize fluorescence rate from the ion of $\sim$20,800 c/s using the PMT. Photons collected via lens B are sent into a single mode fiber (SMF) with a coupling efficiency of $\sim$17$\%$ and sent either directly to an avalanche photon-detector (APD) (Perkin Elmer: SPCM-AQR-15 \cite{disclaimer}) with a QE of $\sim$45$\%$ at 493 nm or sent to the frequency conversion setup (Fig.~\ref{Fig:experiment}(b)) where they are converted to 780 nm and sent to the APD with a specified QE of $\sim$60$\%$ at 780 nm. When sent directly to the APD, the measured 493 nm photon count rate is $\sim$26,600 c/s.

The frequency conversion setup shown in Fig.~\ref{Fig:experiment}(b) spatially combines the 493 nm light and 1343 nm pump using a dichroic mirror. A 20x objective lens then focuses the combined beams onto the PPLN waveguide (Srico: 2000-1004 \cite{disclaimer}). The PPLN output facet couples directly into a 800 nm wavelength polarization maintaining SMF. Light out of the PPLN fiber is free-space coupled to allow for 70 dB of pump filtering and 59 dB of non-converted 493 nm leak-through filtering using a filter set (ThorLabs: FBH780-10 and two Semrock: FF01-1326/SP-25 \cite{disclaimer}). 

\begin{figure}[htbp]
	\centering
	\includegraphics[width=1\columnwidth]{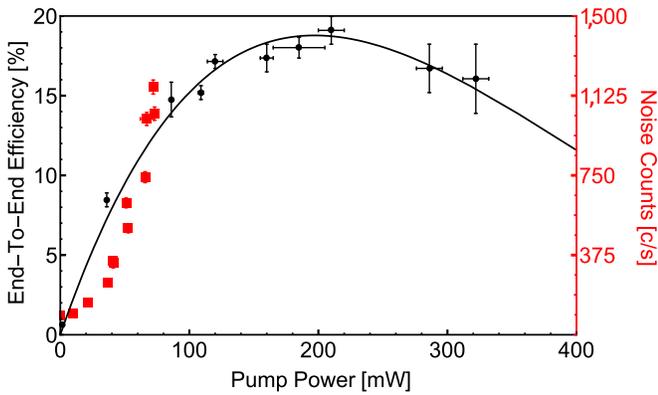}
	\caption{The end-to-end conversion efficiency of the DFG from 493 nm to 780 nm (black circles and left-hand axis) and noise produced in the bandwidth allowed by our filtering (magenta squares and right-hand axis) as a function of pump power coupled into the waveguide. The error bars on the pump power are from fluctuations of the measured power and the error bars on the efficiency are primarily from DFG signal power fluctuations.}
	\label{Fig:DFG}
\end{figure}

To determine the QFC efficiency of the PPLN setup, a 493 nm laser, locked to 607.425690(2) THz, and a pump at 1343 nm, are used as inputs to the PPLN waveguide. The pump power is varied while the DFG optical power is measured after the filters. The end-to-end efficiency is determined by taking the ratio of the measured DFG power to the 493 nm power sent to the device, while correcting for the change in frequency. The end-to-end conversion efficiency depends on the pump power coupled into the waveguide, as shown in Fig.~\ref{Fig:DFG}, with a maximum of  $\sim$19$\%$ at a coupled pump power of $\sim$210 mW. The efficiency is largely limited by the coupling ($\sim$30$\%$) of the 493 nm light into the PPLN.  

While operating at the maximum conversion efficiency increases the number of 780 nm photons produced, the larger pump power required results in increased production of anti-Stokes noise photons \cite{Ballance2015}. Therefore, when frequency converting light from the ion, it is critical to chose a pump power that maximizes the signal-to-noise ratio (SNR) of the DFG light. Figure \ref{Fig:DFG} shows the noise inside the bandwidth of the filters as a function of pump power coupled into the waveguide. We operate the pump at $\sim$40 mW which maximizes the SNR when operating the frequency conversion setup with 493 nm light delivered from the ion.

\begin{figure}[htbp]
	\centering
	\includegraphics[width=1\columnwidth]{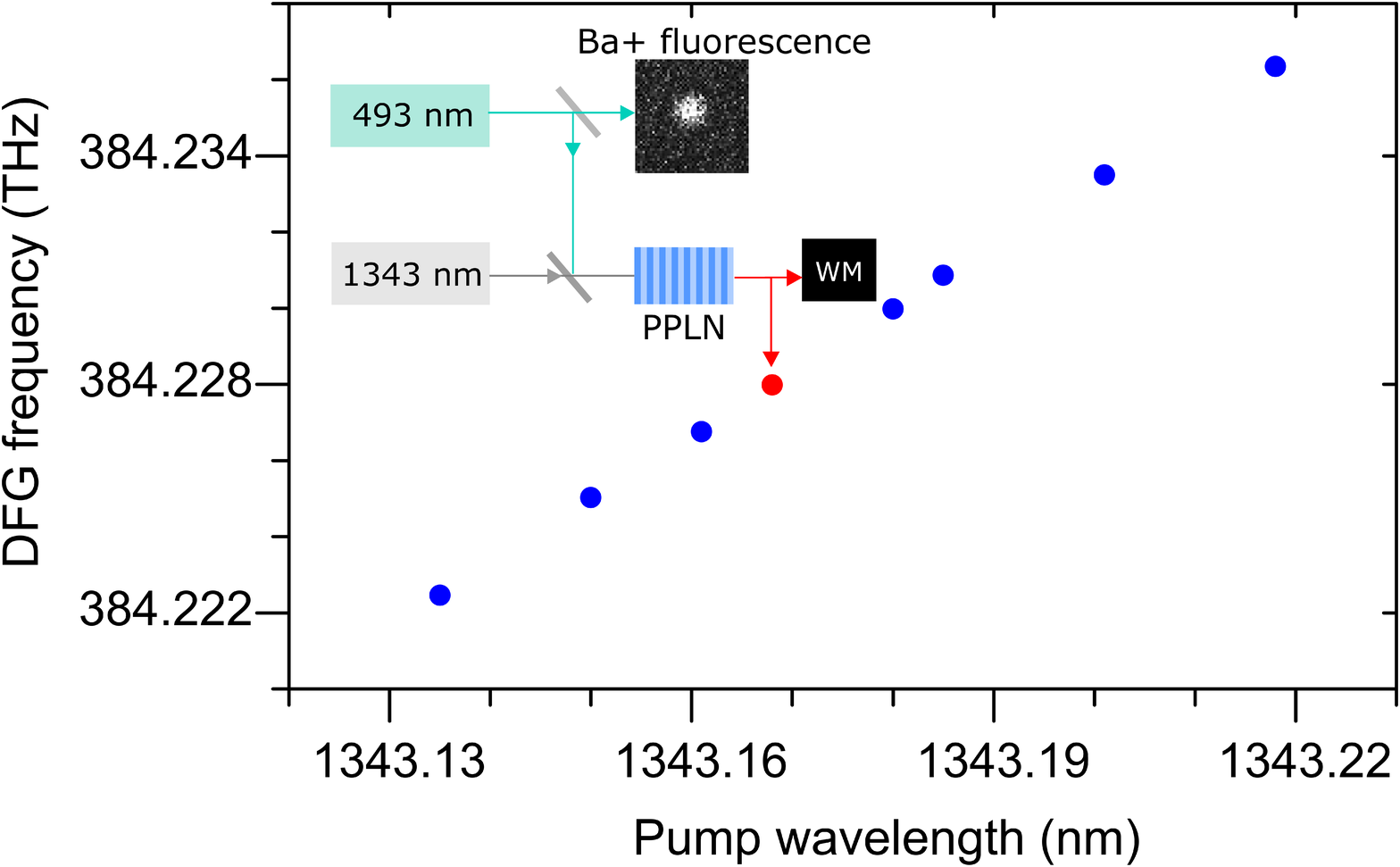}
	\caption{Output DFG output wavelength as a function of the pump wavelength near 1343 nm. The inset is a block diagram of the setup showing simultaneous resonant ion fluorescence when the \Ba ion $^2$S$_{1/2}$ - $^2$P$_{1/2}$ Doppler cooling laser locked to 607.425690(2) THz while the DFG PPLN output is at the $^{87}$Rb $D_2$ resonance of 384.227982(2) THz \cite{Ye1996} (red data point) as measured to a 2 MHz accuracy using a wavemeter (WM) (HighFinesse: WSU/2)\cite{disclaimer}.. The PPLN temperature is held at a constant temperature of 43$^{\circ}$C during this procedure.}
	\label{Fig:tuning curve}
\end{figure}

To facilitate hybrid interactions, it is important that the PPLN DFG output be tunable to resonance with the $^{87}$Rb $D_2$ transition at 384.228 THz \cite{Ye1996} while at the same time the input frequency needs to be resonant with the barium ion. This double resonance condition gives two constraints which are met by identifying appropriate pump wavelength and PPLN oven temperature operation points. To determine these values, the DFG output is maximized at a pump wavelength near 1336 nm at a PPLN oven temperature of 35.5$^{\circ}$C. The DFG's wavelength is then tuned to $^{87}$Rb $D_2$ resonance by increasing the pump wavelength to a value of 1343.169 nm and tuning the PPLN temperature to 43$^{\circ}$C. All of the tuning is carried out with the 493 nm laser light locked to \Ba resonance. Fine tuning of the DFG signal frequency is possible by adjusting the pump wavelength as shown in Fig.~\ref{Fig:tuning curve}. 

For single photon measurements we used the optimal SNR pump power setting, as determined by Fig.~\ref{Fig:DFG} and the pump wavelength and oven temperature setting, as determined by Fig.~\ref{Fig:tuning curve}. With $\sim$40 mW of pump light coupled into the waveguide, we measure a noise level of $\sim$300 c/s on the APD. Blocking light entering the fiber attached to the APD housing results in a dark noise count rate of $\sim$100 c/s. The counts above this noise floor are attributable to leak-through photons in the wavelength range 780 $\pm$ 5 nm produced by anti-Stokes Raman photons from the pump. To show this noise is not leakage of the pump through the filters, 40 mW of pump light was sent directly through the filter set, bypassing the PPLN, and the same 100 c/s dark noise rate was observed. Importantly, there were no detectable levels of 493 nm or 650 nm ion-light making it through the filters by observing APD count rates equal to that of the dark counts when only the ion-light is present. Therefore, any sub-unitary second-order intensity correlation measurements detected between events on the APD and the PMT can only be caused by correlations between frequency converted 780 nm photons and unconverted 493 nm photons. 

When both 493 nm photons from the ion and the 40 mW pump light are present in the PPLN waveguide, the measured count rate of the the 780 nm DFG light is $\sim$600 c/s, giving a SNR of $\sim$2. A conversion efficiency $\sim$2.3 $\%$, including all optical and fiber coupling losses, is determined by taking the ratio of the converted photon count rate to the 493 nm photon count rate as measured directly on the APD ($\approx$ 26,100 c/s). Taking into account a factor of two loss in a fiber patch cable between the ion and QFC tables, another factor of two loss for polarization selectivity of the QFC process and other optical loses, this value is consistent with the classically measured value shown in Fig.~\ref{Fig:DFG}.

The measurement of the g\textsuperscript{(2)}($\tau$) correlation function of single photons produced by the Ba\textsuperscript{+} ion was performed using the light collected from two sides of the ion, as a function of the relative delay, $\tau$, between detection events on either side. This approach is equivalent to a Hanbury-Brown-Twiss apparatus where the ion can be treated as the beam splitter. The photons were collected while continuously Doppler cooling the ion. The measurement was performed first between non-frequency converted 493 nm photons emitted by the ion and, then, between non-frequency converted photons and frequency converted 780 nm photons. A PMT was used to detect 493 nm photons collected by lens A and these detection events were correlated to APD detected photons collected by lens B, either directly from the ion or via the frequency conversion setup (shown in Fig.~\ref{Fig:experiment}(a)). 

The output of each detector is sent to a time-correlated two-channel single photon counter (TCSPC) with a resolution of 512 ps (PicoQuant: PicoHarp 300). Using the counter's histogram mode, the arrival of a pulse from the PMT triggers a timing circuit which measures the time until a pulse is received from the APD. An electrical delay line was added to the PMT line to allow measurement of negative delay times. The time between pulses on each detector was then binned into 512 ps wide bins. To calculate the g\textsuperscript{(2)}($\tau$), the raw coincidence counts were normalized using \cite{munoz2016,Brouri00}

\begin{equation}
g^{(2)}(\tau)=\frac{N(\tau)}{N_{PMT} N_{APD} \delta_t T}.
\label{gtwo}
\end{equation}

\begin{figure}[htbp]
	\centering
	\includegraphics[width=1\columnwidth]{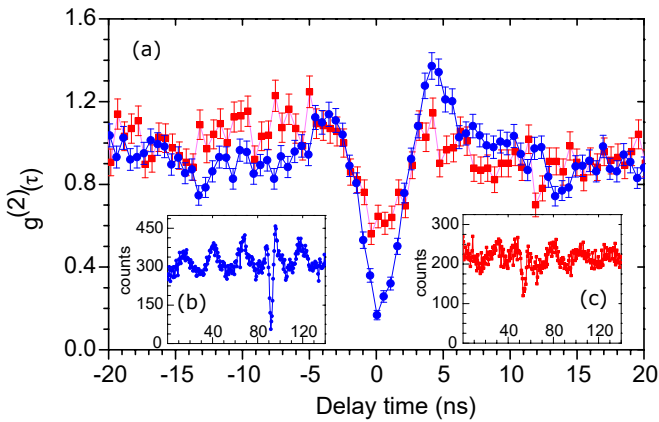}
	\caption{(a) The g\textsuperscript{(2)}($\tau$) correlation function measured between non-frequency converted 493 nm photons (blue) where g\textsuperscript{(2)}(0) = $0.167 \pm .022$ and between 493 nm and converted 780 nm photons (magenta) where g\textsuperscript{(2)}(0) = $0.645 \pm .055$. These measurements agree with expected values given by Eqn.~\ref{eq:gexpected} given that $\delta_t=$ 512 ps. The error bars represent the shot noise from each detector. The insets show the raw, un-normalised, coincidence counts where (b) $N_{APD}=26,600$ c/s and $N_{PMT}=19,700$ c/s and $T=$ 20 minutes and for the (c) $N_{APD}=930$ c/s and $N_{PMT}=20,500$ c/s and $T=$ 6.13 hours.}
	\label{Fig:g2}
\end{figure}

Here, $N(\tau)$ is the number of coincidence counts in a bin, $\delta_t$ is the size of the time bin (512 ps), $T$ is the total integration time, and $N_{PMT}$ and $N_{APD}$ are the total count rates in the PMT and APD channels respectively. The normalized and delay-line compensated $g^{(2)}(\tau)$ measured between non-converted photons (blue) and converted and non-converted photons (magenta) are shown in Fig.~\ref{Fig:g2}. The insets show the raw coincidence count data with the electronic delays of 91.65 ns and 53.76 ns present. The periodic oscillation of the counts is due to micromotion of the ion in the trap at a frequency of $\Omega/ 2\pi$ = 38.4 MHz. The measured g\textsuperscript{(2)}(0) of 493 nm photons directly emitted by the ion and between 493 nm photons and frequency converted 780 nm photons were measured to be 0.167 $\pm$ 0.022 and 0.645 $\pm$ 0.055 respectively.   

In a noiseless $g^{(2)}{(\tau)}$ measurement with infinitely small bin widths, zero coincident counts are expected at $\tau = 0$ from a single photon source, such as an ion. Experimentally however, a finite bin width must be used and noise detector clicks are inevitable. The finite bin width is used to calculate a minimum $g^{(2)}(0)$ of $a = 0.035$, using the methods described in \cite{munoz2016}. To take into account the effects of noise and finite bin width, the following is used \cite{Becher2001}

\begin{equation}
g^{(2)}_{exp}(t) = 1 + \left( \frac{SNR}{1+SNR}\right)^{2} (a - 1),
\label{eq:gexpected}
\end{equation}

\noindent
where $SNR$ is the ratio of the histogram counts per second due to the signal from the ion to the histogram counts per second due to correlations with noise. The latter includes PMT noise-APD noise correlations, PMT signal-APD noise correlations and PMT noise-APD signal correlations.

For the correlation measurement solely between the unconverted 493 nm photons from the ion, the main noise sources were from correlations between the PMT signal and the APD noise and between the PMT noise and APD signal. The histogram count rates from each of these sources was measured to be ~276 c/s and ~359 c/s respectively. Including a small PMT noise-APD noise contribution (6 c/s) a total noise rate of ~641 c/s was measured. Given a total histogram count rate of 10,800 c/s a SNR of ~15.8 is achieved. The expected $g^{(2)}_{exp}(0)$ is therefore found to be 0.146. This value is in agreement with the measured value of $0.167 \pm .022$.

The measurement between the unconverted 493 nm photons and converted 780 nm photons has noise counts which are dominated by correlations between signal PMT counts and APD noise counts. The measured histogram count rate due to such correlations was found to be 200 c/s and the other noise sources produced a combined 20 c/s. With a total count rate of 615 c/s, this corresponds to a signal to noise ratio of 1.80. The measured value of $g^{(2)}(0)$ was found to be $0.645 \pm 0.055$ in agreement with the expected value of $0.602$, and is also significantly below the classical limit, demonstrating the conversion of single photons from 493 nm to 780 nm.

During the course of this work, photons from a Ca$^+$ ion were frequency converted to 1530 nm preserving their quantum statistics \cite{Keller17} and, separately, were frequency converted to 1310 nm with ion-photon entanglement demonstrated \cite{Eschner17}. 

In conclusion, we demonstrate the conversion of photons from a single trapped ion into optical frequencies which are resonant with a neutral atomic system. The conversion of photons emitted from a trapped ion into the near-infrared provides the ability to network ions over significantly larger distances that is achievable with their native wavelengths due to decreased losses in optical fibers. This result paves the way for inter-ion and also hybrid species quantum networking.

\vspace{10mm}
\noindent     
\textbf{Acknowledgments}
\begin{acknowledgments}
This work is supported by the Center for Distributed Quantum Information (CDQI) and the Army Research Laboratory Open Campus Initiative. 
\end{acknowledgments}

\vspace{5mm}
\noindent
\textbf{Author Contributions}
\begin{acknowledgments}
J.D.S., J.H. and Q.Q. all contributed contributed to the design, construction and carrying out of the experiment. All authors contributed to this manuscript. 
\end{acknowledgments}


\bibliography{ion_qfc}

\end{document}